\documentclass[amssymb,twocolumn,eqsecnum,aps]{revtex4}
\usepackage{amsmath}
\usepackage{enumerate}
\usepackage{graphicx}
\DeclareGraphicsExtensions{.pdf,.png,.jpg,.eps}
\begin{document}
\def \bnab {{\bf\nabla}\hskip-8.8pt{\bf\nabla}\hskip-9.1pt{\bf\nabla}}
\title{Continuum Electrodynamics of a Piecewise-Homogeneous Linear Medium}
\author{Michael E. Crenshaw}
\affiliation{US Army Aviation and Missile Research, Development, and Engineering Center, Redstone Arsenal, AL 35898, USA}
\date{\today}
\begin{abstract}
The energy--momentum tensor and the tensor continuity equation serve as
the conservation laws of energy, linear momentum, and angular momentum
for a continuous flow.
Previously, we derived equations of motion for macroscopic
electromagnetic fields in a homogeneous linear dielectric medium that
is draped with a gradient-index antireflection coating
(J. Math Phys. {\bf 55,} 042901 (2014) ).
These results are consistent with the electromagnetic tensor continuity
equation in the limit that reflections and the accompanying surface
forces are negligible thereby satisfying the condition of an unimpeded
flow in a thermodynamically closed system.
Here, we take the next step and derive equations of motion for the
macroscopic fields in the limiting case of a piecewise-homogeneous
simple linear dielectric medium.
The presence of radiation surface forces on the interface between two
different homogeneous linear materials means that the energy--momentum
formalism must be modified to treat separate homogeneous media in which
the fields are connected by boundary conditions at the interfaces.
We demonstrate the explicit separation of the total momentum into a
field component and a material motion component, we derive the radiation
pressure that transfers momentum from the field to the material, we
derive the electromagnetic continuity equations for a piecewise
homogeneous dielectric, and we provide a lucid reinterpretation of
the Jones and Richards experiment.
\end{abstract}
\maketitle
\par
\section{Introduction}
\par
The energy--momentum tensor is an innate and compelling aspect of
energy and momentum conservation in a continuous flow \cite{BIFoxMcD}.
Recently \cite{BIy,BIz,BISPIE,BIx}, we used global conservation
principles to construct the total energy--momentum tensor for a
thermodynamically closed system consisting of a quasimonochromatic
optical pulse and a homogeneous simple linear medium that is draped with
a gradient-index antireflection coating.
Regarding the total energy--momentum tensor and the tensor continuity
equation as fundamental, we derived equations of motion for the
macroscopic fields.
The formulation of continuum electrodynamics that was derived in our
previous work \cite{BIy,BIz,BISPIE,BIx,BIf,BILag} was limited to
homogeneous materials with a gradient-index antireflection coating.
In this article, we develop the theory of continuum electrodynamics for
the more usual situation of a piecewise-homogeneous linear dielectric
medium.
One of the major differences with the Maxwell theory is that the Fresnel
relations can no longer be derived from the application of Stoke's
theorem to the Faraday and Maxwell--Amp\`ere Laws.
Instead, we derive the Fresnel relations from the electromagnetic wave
equation and conservation of energy.
We obtain the field and material components of the total momentum,
derive the radiation pressure, and derive the electromagnetic
continuity equations for a piecewise homogeneous medium.
We provide an interpretation of the Jones and
Richards \cite{BIExp} measurement of the optical force on a mirror
immersed in a dielectric fluid.
\par
\section{Equations of Motion}
\par
We take as a given that propagation of the electromagnetic field is
characterized by the wave equation
\begin{equation}
\nabla\times(\nabla\times{\bf A})
+\frac{n^2}{c^2}\frac{\partial^2{\bf A}}{\partial t^2}=0
\label{EQxxx2.01}
\end{equation}
in a limit in which absorption can be neglected.
Dispersion is treated parametrically for the arbitrarily long
quasimonochromatic fields that are considered here.
The electromagnetic wave equation, Eq.~(\ref{EQxxx2.01}), is a mixed
second-order differential equation that can be written in terms of
first-order differential equations.
To that end, we define the macroscopic magnetic field
\begin{equation}
{\bf B}=\nabla\times{\bf A}
\label{EQxxx2.02}
\end{equation}
and a second macroscopic field
\begin{equation}
{\bf \Pi}=\frac{n}{c}\frac{\partial{\bf A}}{\partial t} \,.
\label{EQxxx2.03}
\end{equation}
A Maxwell--Amp\`ere-like law
\begin{equation}
\nabla\times{\bf B}+\frac{n}{c}\frac{\partial{\bf \Pi}}{\partial t}=0 
\label{EQxxx2.04}
\end{equation}
results from the substitution of the definitions of the macroscopic
fields, Eqs.~(\ref{EQxxx2.02}) and (\ref{EQxxx2.03}), into the wave
equation, Eq.~(\ref{EQxxx2.01}).
Two more equations of motion are derived from the definitions
of the fields.
Thompson's Law
\begin{equation}
\nabla\cdot{\bf B}=0
\label{EQxxx2.05}
\end{equation}
is obtained from the divergence of Eq.~(\ref{EQxxx2.02}) with the vector
identity that the divergence of the curl of any vector is zero.
The curl of Eq.~(\ref{EQxxx2.03})
\begin{equation}
\nabla\times{\bf \Pi}-\frac{n}{c}\frac{\partial{\bf B}}{\partial t}
=\frac{\nabla n}{n}\times{\bf \Pi}
\label{EQxxx2.06}
\end{equation}
is our variant of the Faraday Law.
Taking the divergence of Eq.~(\ref{EQxxx2.04}) and integrating with
respect to time, we obtain the Gauss-like law
\begin{equation}
\nabla\cdot{\bf \Pi}=-\frac{\nabla n}{n}\cdot{\bf \Pi} \,.
\label{EQxxx2.07}
\end{equation}
A constant of integration has been suppressed in the absence of
charges.
Note that each field equation is algebraically equivalent to its
counterpart in the macroscopic Maxwell equations
\begin{equation}
\nabla\times{\bf B}-\frac{n^2}{c}\frac{\partial{\bf E}}{\partial t}=0
\label{EQxxx2.08}
\end{equation}
\begin{equation}
\nabla\cdot{\bf B}=0
\label{EQxxx2.09}
\end{equation}
\begin{equation}
\nabla\times{\bf E}+\frac{1}{c}\frac{\partial{\bf B}}{\partial t}=0
\label{EQxxx2.10}
\end{equation}
\begin{equation}
\nabla\cdot (n^2{\bf E})=0 
\label{EQxxx2.11}
\end{equation}
if we define ${\bf \Pi}=-n{\bf E}$, which we have every right to do
under the auspices of Maxwellian continuum electrodynamics.
However, the different sets of motional equations for macroscopic fields
have different tensorial and relativistic properties.
This means that Maxwellian continuum electrodynamics, which is
fundamentally a vector theory, admits improper tensor transformations
of coordinates for the coupled equations \cite{BILag}.
The derivation of the motional equations
Eqs.~(\ref{EQxxx2.04})--(\ref{EQxxx2.07})
from Lagrangian field theory appears elsewhere \cite{BILag}.
Now, it can be argued that Eqs.~(\ref{EQxxx2.04})--(\ref{EQxxx2.07})
violate relativity because the Lorentz factor is
$\gamma_d=1/\sqrt{1-n^2v^2/c^2}$ instead of
$\gamma_v=1/\sqrt{1-v^2/c^2}$.
However, that argument presupposes transformations between a coordinate
system in the dielectric and a coordinate system in a vacuum laboratory
(lab) frame.
If we consider, instead, transformations between two coordinate systems
in an arbitrarily large region of space in which the speed of light is
$c/n$ \cite{BIRosen}, as we should, then $\gamma_d$ is the correct
Lorentz fator and Eqs.~(\ref{EQxxx2.04})--(\ref{EQxxx2.07}) are,
in fact, compatible with relativity \cite{BILag}.
\par
Returning to the equations of motion for the macroscopic fields,
Eqs.~(\ref{EQxxx2.04})--(\ref{EQxxx2.07}), we can identify two limiting
cases of particular interest.
The first is when the gradient of the index of refraction $\nabla n$
is sufficiently small that reflections and Helmholtz forces can be
neglected \cite{BIy}.
It is in this limit of an unimpeded continuous flow of electromagnetic
radiation that the energy and momentum densities can be expressed
through continuity equations and a total energy--momentum tensor 
for a thermodynamically closed system.
This case was treated in Refs.~ \cite{BIy,BIz,BISPIE,BIx,BIf,BILag}.
Here, we consider a quasimonochromatic optical pulse that passes from
one simple linear medium into a second such medium through a planar
interface at normal incidence.
In this limit of piecewise-homogeneous media, we need only retain the
homogeneous parts of Eqs.~(\ref{EQxxx2.04})--(\ref{EQxxx2.07}) to obtain
\begin{equation}
\nabla\times{\bf B}+\frac{n}{c}\frac{\partial{\bf \Pi}}{\partial t}=0
\label{EQxxx2.12}
\end{equation}
\begin{equation}
\nabla\cdot{\bf B}=0
\label{EQxxx2.13}
\end{equation}
\begin{equation}
\nabla\times{\bf \Pi}-\frac{n}{c}\frac{\partial{\bf B}}{\partial t}=0
\label{EQxxx2.14}
\end{equation}
\begin{equation}
\nabla\cdot{\bf \Pi}=0 \, ,
\label{EQxxx2.15}
\end{equation}
where the fields are connected by boundary conditions at the interface
between different homogeneous linear materials.
While the derivation of the equations of motion for macroscopic fields
in a homogeneous medium from Eqs.~(\ref{EQxxx2.04})--(\ref{EQxxx2.07})
is obvious, the usage of Eqs.~(\ref{EQxxx2.12})--(\ref{EQxxx2.15}) for
piecewise-homogeneous matter remains to be investigated.
\par
\section{The Fresnel Relations}
\par
First, we demonstrate how our results apply to the most obvious
issue relating to piecewise-homogeneous matter, the Fresnel relations.
Because the equations of motion for macroscopic fields in a
piecewise-homogeneous linear medium have changed, there is a question
that arises as to how the Fresnel relations survive.
Applying Stokes's theorem to the macroscopic Maxwell curl equations,
Eqs.~(\ref{EQxxx2.08}) and (\ref{EQxxx2.10}), it was long ago found that
the tangential components of the electric field ${\bf E}$ and the
magnetic auxiliary field ${\bf H}={\bf B}$ are continuous at a planar
interface between linear homogeneous dielectrics.
Similarly, the normal components of the displacement field
${\bf D}=n^2{\bf E}$ and the magnetic field ${\bf B}$ were shown to be
continuous by the divergence theorem applied to the Maxwell divergence
equations, Eqs.~(\ref{EQxxx2.09}) and (\ref{EQxxx2.11}).
The simultaneous continuity of the transverse ${\bf E}$ and ${\bf H}$
fields and the normal parts of the ${\bf D}$ and ${\bf B}$ fields leads
to the Fresnel relations.
Because Eqs.~(\ref{EQxxx2.06}) and (\ref{EQxxx2.07}) are inhomogeneous,
the transverse and normal components of the macroscopic field
${\bf \Pi}$ are not continuous at the material interface, even though
the appearance of Eqs.~(\ref{EQxxx2.14}) and (\ref{EQxxx2.15}) might
seem to suggest otherwise.
\par
We treat the propagation of a quasimonochromatic optical pulse through
a piecewise-homogeneous simple linear medium.
The material is initially stationary in the laboratory frame of
reference and is rigidly attached to a support.
We take as a given that propagation of the electromagnetic field is
characterized by the wave equation
\begin{equation}
\nabla\times(\nabla\times{\bf A})
+\frac{n^2({\bf r})}{c^2}\frac{\partial^2{\bf A}}{\partial t^2}=0 \,.
\label{EQxxx3.01}
\end{equation}
The pulse is sufficiently monochromatic that dispersion can be neglected
in accordance with the characterization of a simple linear medium.
Further, the simple linear material with its support is assumed to be
arbitrarily massive so that $n({\bf r})$ can be treated as a
time-independent function of space in the laboratory frame of reference.
In order to not overly complicate matters, we assume normal incidence
and adopt the plane-wave limit for the field.
The vector potential, in the plane-wave limit,
$$
{\bf A}(z,t)=\frac{1}{2}\left (
{\bf\tilde A}(z,t)e^{-i(\omega_d t-k_d z)}+
{\bf\tilde A}^*(z,t)e^{i(\omega_d t-k_d z)}
\right )
$$
can be written in terms of an envelope function $\tilde A(z,t)$ and a
carrier wave with center frequency $\omega_d$.
Here, $k_d$ is the amplitude of the wave vector
${\bf k}_d=(n\omega_d /c){\bf\hat e}_z$ that is associated with the
center frequency of the field, and ${\bf\hat e}_z$ is a unit vector in
the direction of propagation along the $z$ axis.
The vector potential amplitude ${\bf\tilde A}$ is not slowly varying if
there is a backward propagating field component.
Figure 1 shows a one-dimensional representation of the 
amplitude of the incident field
$\tilde A_0(z)=
\left ( {\bf\tilde A}(z,t_0)\cdot{\bf\tilde A}^*(z,t_0) \right )^{1/2}$
about to enter a simple linear medium with $n_2=1.40$ at normal
incidence from the vacuum $n_1=1$.
\begin{figure}
\includegraphics[scale=0.70]{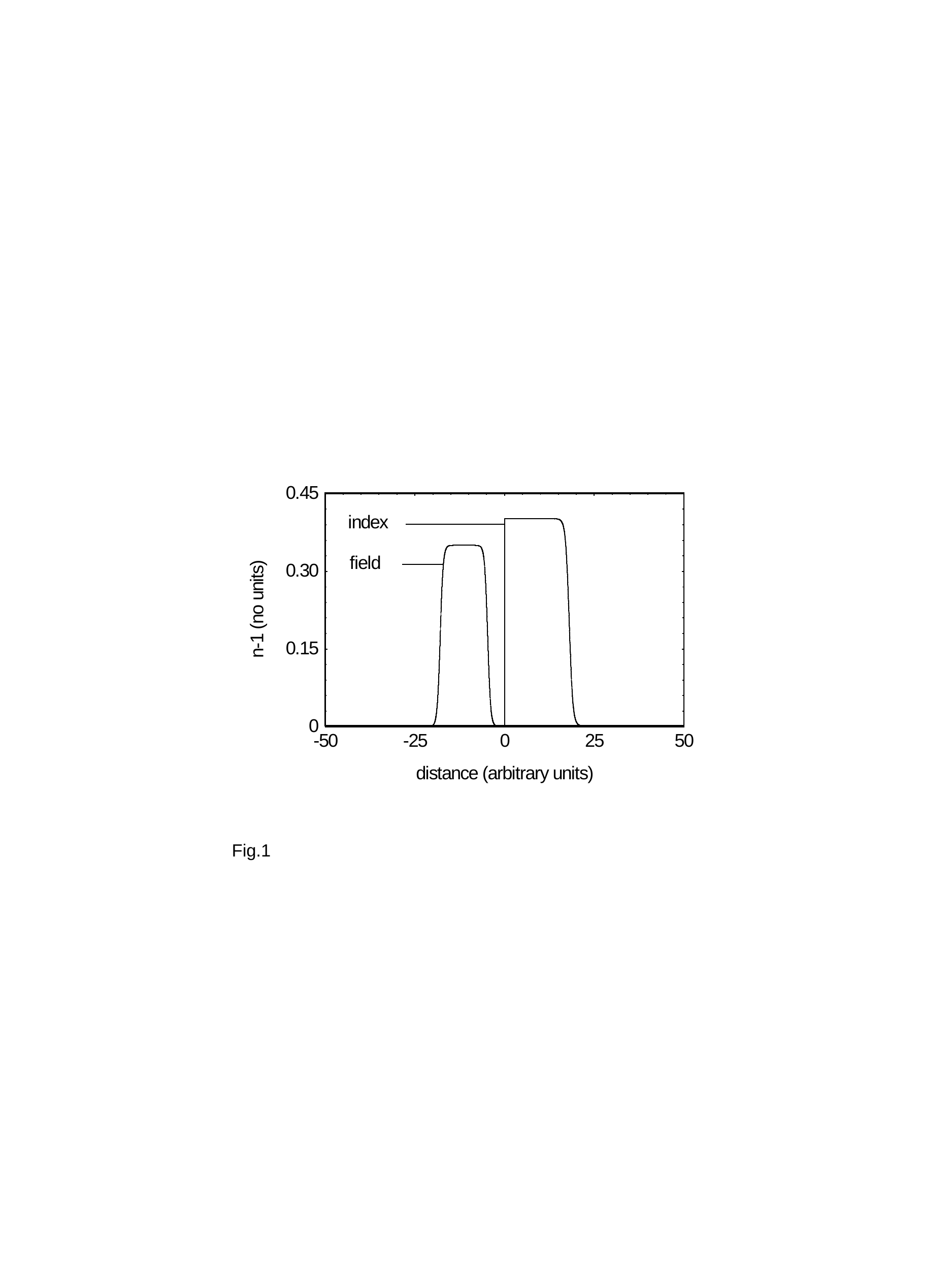}
\caption{Amplitude of the incident field envelope (arb. units)}
\label{fig1}
\end{figure}
Figure 2 presents a time-domain numerical solution of the wave equation
at a later time $t_1$ depicted by
$\tilde A_1(z)=
\left ({\bf\tilde A}(z,t_1)\cdot{\bf\tilde A}^*(z,t_1)\right )^{1/2}$\,.
The reflected field and the refracted field have separated and the
refracted field is entirely inside the medium.
The refracted pulse has not propagated as far as it would have
propagated in the vacuum due to the reduced speed of light $c/n$ in the
material.
In addition, the spatial extent of the refracted pulse in the medium
is $w_t=n_1 w_i/n_2$ in terms of the width $w_i$ of the incident pulse.
\begin{figure}
\includegraphics[scale=0.70]{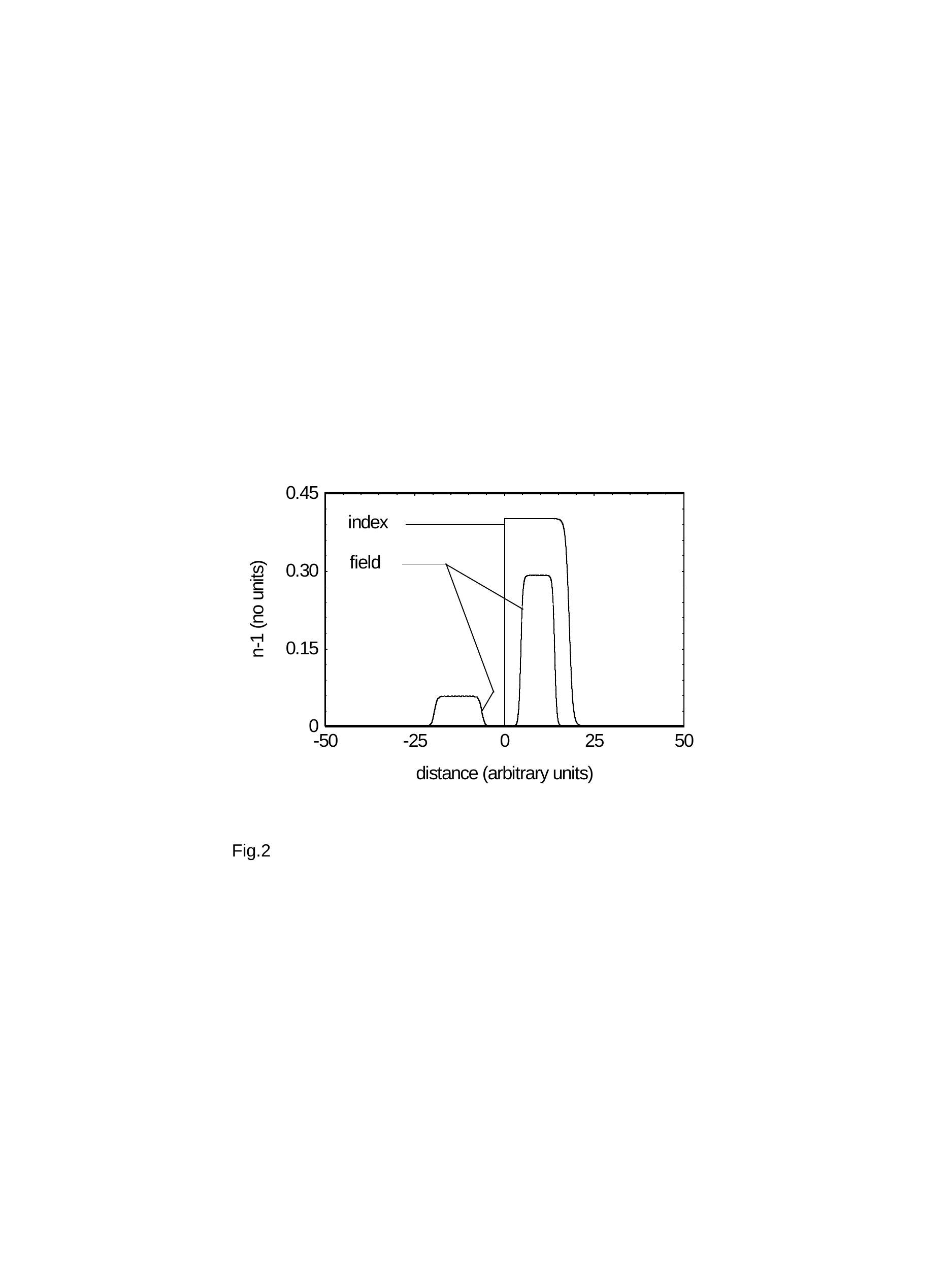}
\caption{Refracted field entirely within the linear medium and separated
from the reflected field.}
\label{fig2}
\end{figure}
As shown in Fig. 2, the amplitudes of the reflected and refracted fields
are different from the amplitude of the incident field.
\par
We would like a formula for determining how much of the incident pulse
goes into the reflected field and how much is refracted.
Although the Fresnel relations are the necessary formulas, their use is
problematic at this point because their provenance from the macroscopic
Maxwell equations, Eqs.~(\ref{EQxxx2.08})--(\ref{EQxxx2.11}), is suspect.
Because the macroscopic field ${\bf \Pi}$ is not continuous at a step
index boundary, we present a derivation of the Fresnel relations that is
based on the wave equation and conservation of energy.
Applying Stokes's theorem to the wave equation, Eq.~(\ref{EQxxx3.01}),
we have
\begin{equation}
\oint_C (\nabla\times{\bf A}) \cdot d{\bf l}=
\int_S (\nabla\times(\nabla \times{\bf A}))\cdot{\bf\hat n} \, da \, .
\label{EQxxx3.02}
\end{equation}
Consider a thin right rectangular box or ``Gaussian pillbox'' that
straddles the interface between the two mediums with the large surfaces
parallel to the interface.
Then $S$ is the surface of the pillbox,
$da$ is an element of area on the surface, and ${\bf\hat n}$
is an outwardly directed unit vector normal to $da$.
There is no contribution to the surface integral from the large surfaces
for our normally incident field because
$\nabla\times(\nabla\times{\bf A})$ is orthogonal to ${\bf\hat n}$.
The contributions from the smaller surfaces can be neglected as the box
becomes arbitrarily thin. 
Then
\begin{equation}
\oint_C (\nabla\times{\bf A}) \cdot d{\bf l}= 0 \, .
\label{EQxxx3.03}
\end{equation}
We choose the closed contour $C$ in the form of a rectangular Stokesian
loop with sides that bisect the two large surfaces and two of the small
surfaces on opposite sides of the pillbox.
Here, $d{\bf l}$ is a directed line element that lies on the
contour, $C$.
Then $C$, like $S$, straddles the material interface.
For normal incidence in the plane-wave limit, the field
$\nabla\times{\bf A}$ can be oriented along the long sides of the 
contour $C$.
Performing the contour integration in Eq.~(\ref{EQxxx3.03}), the
contribution from the short sides of the contour are neglected as the
loop is made vanishingly thin and we obtain
\begin{equation}
(\nabla\times{\bf A})_1\cdot\Delta {\bf l}_1 +
(\nabla\times{\bf A})_2\cdot\Delta {\bf l}_2=0
\label{EQxxx3.04}
\end{equation}
from the long sides, 1 and 2, of the contour.
\par
For linearly polarized radiation, we can write the
vector potential of the incident, reflected, refracted, and 
transmitted waves as
\begin{equation}
{\bf A}_i={\bf\hat e}_x \tilde A_i e^{-i(\omega_d t-k_1z)}
\label{EQxxx3.05}
\end{equation}
\begin{equation}
{\bf A}_r={\bf\hat e}_x \tilde A_r e^{-i(\omega_d t+k_1z)}
\label{EQxxx3.06}
\end{equation}
\begin{equation}
{\bf A}_t={\bf\hat e}_x \tilde A_t e^{-i(\omega_d t-k_2z)} 
\label{EQxxx3.07}
\end{equation}
\begin{equation}
{\bf A}_T={\bf\hat e}_x \tilde A_T e^{-i(\omega_d t-k_1z)} \, .
\label{EQxxx3.08}
\end{equation}
where $k_1=n_1\omega_d/c$, $k_2=n_2\omega_d/c$, and
${\bf\hat e}_x$ is a unit polarization vector.
It is understood that we use the real part of complex fields
and neglect double frequency terms in field products.
For convenience, the scalar amplitudes are taken to be real.
Using the fact that the line elements $\Delta{\bf l}_1$ and
$\Delta{\bf l}_2$ in Eq.~(\ref{EQxxx3.04}) are equal and opposite,
we obtain a relation
\begin{equation}
n_1(\tilde A_i-\tilde A_r)=n_2\tilde A_t
\label{EQxxx3.09}
\end{equation}
between the amplitudes of the incident, reflected, and refracted
fields.
In order to derive boundary conditions, we need another such relation.
\par
For a stationary simple linear material, the electromagnetic energy
\begin{equation}
U=\int_{\sigma}\frac{1}{2}\left (
\frac{n^2}{c^2}\left (\frac{\partial{\bf A}}{\partial t}\right )^2
+(\nabla\times{\bf A})^2
\right ) dv
\label{EQxxx3.10}
\end{equation}
is conserved.
Here, the volume of integration, which includes all fields present,
has been extended to all-space $\sigma$.
The total energy is invariant in time by virtue of being conserved.
The total energy at time $t_0$, the incident energy $U(t_0)=U_i$,
is equal to the total energy at a later time $t_1$, $U(t_1)=U_r+U_t$,
which is the sum of the reflected energy $U_r$ and the refracted
energy $U_t$ when the refracted field is entirely within the medium.
\par
In terms of the incident, reflected, and transmitted energy, the energy
balance $U(t_0)=U(t_1)$ is 
\begin{equation}
U_i=U_r+U_t \, .
\label{EQxxx3.11}
\end{equation}
Substituting Eqs.~(\ref{EQxxx3.05})--(\ref{EQxxx3.07}) into the formula
for the energy, Eq.~(\ref{EQxxx3.10}), and expressing the energy 
balance, Eq.~(\ref{EQxxx3.11}), in
terms of the amplitudes of the incident, reflected, and transmitted
vector potential results in
\begin{equation}
\int_{\sigma} n_1^2 \tilde A_i^2 dv=
\int_{\sigma} n_1^2 \tilde A_r^2 dv+
\int_{\sigma} n_2^2 \tilde A_t^2 dv \, .
\label{EQxxx3.12}
\end{equation}
In order to facilitate the integration of Eq.~(\ref{EQxxx3.12}), we
choose the incident pulse to be rectangular with a nominal width
of $w_i$.
The pulse has a finite rise time and a finite fall time to reduce
ringing, but the short transition region can be neglected compared
to the arbitrarily large width of the pulse.
The refracted pulse has a width of $n_1w_i/n_2$ due to the change in the
velocity of light between the two media.
Then, evaluating the integrals of Eq.~(\ref{EQxxx3.12}) results in
\begin{equation}
n_1^2 \tilde A_i^2=n_1^2\tilde A_r^2+n_2n_1 \tilde A_t^2 \,.
\label{EQxxx3.13}
\end{equation}
Grouping terms of like refractive index, the previous equation
\begin{equation}
n_1 \left (\tilde A_i^2 - \tilde A_r^2 \right ) = n_2 \tilde A_t^2 
\label{EQxxx3.14}
\end{equation}
becomes more suggestive as
\begin{equation}
n_1 \left (\tilde A_i - \tilde A_r \right )
\left (\tilde A_i + \tilde A_r \right )
 = n_2 \tilde A_t^2
\label{EQxxx3.15}
\end{equation}
by factoring the binomial.
The second-order equation can be written as two first-order equations.
Substituting Eq.~(\ref{EQxxx3.09}) into Eq.~(\ref{EQxxx3.15}), we have
the unique decomposition
\begin{equation}
n_1 \left (\tilde A_i - \tilde A_r \right ) = n_2 \tilde A_t
\label{EQxxx3.16}
\end{equation}
\begin{equation}
\left (\tilde A_i + \tilde A_r \right ) = \tilde A_t 
\label{EQxxx3.17}
\end{equation}
of Eq.~(\ref{EQxxx3.13}).
We eliminate $\tilde A_t$ from Eq.~(\ref{EQxxx3.16}) using
Eq.~(\ref{EQxxx3.17}) to obtain
\begin{equation}
\frac{\tilde A_r}{\tilde A_i}=\frac{n_1-n_2}{n_1+n_2} \,.
\label{EQxxx3.18}
\end{equation}
Subsequently, we eliminate $\tilde A_r$ to get
\begin{equation}
\frac{\tilde A_t}{\tilde A_i}=\frac{2n_1}{n_1+n_2} \,.
\label{EQxxx3.19}
\end{equation}
We see that the usual Fresnel relations can be derived without invoking 
Maxwell's equations.
Conservation of energy, by itself, is sufficient to derive
Eq.~(\ref{EQxxx3.13}).
However, there are several ways that Eq.~(\ref{EQxxx3.13}) can be
decomposed into two first-order equations.
The application of Stoke's theorem to the wave equation guarantees
uniqueness of the decomposition represented by Eqs.~(\ref{EQxxx3.16})
and (\ref{EQxxx3.17}) and the Fresnel relations,
Eqs.~(\ref{EQxxx3.18}) and (\ref{EQxxx3.19}).
\par
\section{Momentum Conservation in Piecewise Homogeneous Media}
\par
The correct form for the momentum of the electromagnetic field in a
dielectric is the subject of the century-old Abraham--Minkowski
controversy \cite{BIPfei,BIMilBoy,BIKemp,BIBaxL,BIBarL,BIMan,BIGriffRL}.
The currently accepted resolution of the controversy, due to
M\o ller \cite{BIMol}, Penfield and Haus \cite{BIPenHau},
Pfeifer {\it et al.} \cite{BIPfei}, and others,
is that the issue is undecidable because neither the Abraham momentum
nor the Minkowski momentum is the total momentum.
If we adopt this viewpoint, then the Abraham momentum and the Minkowski
momentum are irrelevant.
Here we use conservation of total energy and conservation of total
momentum to show that each component of the total linear momentum,
reflected, refracted or transmitted, and kinematic, has a definite
expression in terms of the macroscopic fields.
\par
The energy and momentum of an electromagnetic pulse in the vacuum
\begin{equation}
U_v=\int_{\sigma}\frac{1}{2}
\left ( {\bf \Pi}_v^2+{\bf B}_v^2 \right )dv
\label{EQxxx4.01}
\end{equation}
\begin{equation}
{\bf G}_v=\int_{\sigma}\frac{{\bf B}_v\times{\bf \Pi}_v}{c} dv
\label{EQxxx4.02}
\end{equation}
are well-defined and settled.
Here, $v$ denotes a quantity that is based in the vacuum, $n_1=1$.
The total energy and the total momentum of our system in the
initial configuration at $t_0$ as shown in Fig.~1 are considered to be
given quantities.
Figure 3 shows the result of continuing the numerical solution of 
the wave equation, Eq.~(\ref{EQxxx3.01}), until the pulse has propagated
completely through the medium.
The incident, reflected, and transmitted fields are in vacuum with
well-defined energies so that we can write an energy balance equation
$U_i=U_r+U_T+U_{kinematic}$.
Here, $U_{kinematic}$ is the kinematic energy of a solid block of
dielectric material.
Writing the components of the energy balance equation in terms of the
corresponding vector potential amplitudes, we have
\begin{equation}
\int_{\sigma}\frac{\omega_d^2}{2c^2} \tilde A_i^2 dv =
\int_{\sigma}\frac{\omega_d^2}{2c^2} \tilde A_r^2 dv+
\int_{\sigma}\frac{\omega_d^2}{2c^2} \tilde A_T^2 dv 
+U_{kinematic}\,.
\label{EQxxx4.03}
\end{equation}
\begin{figure}
\includegraphics[scale=0.70]{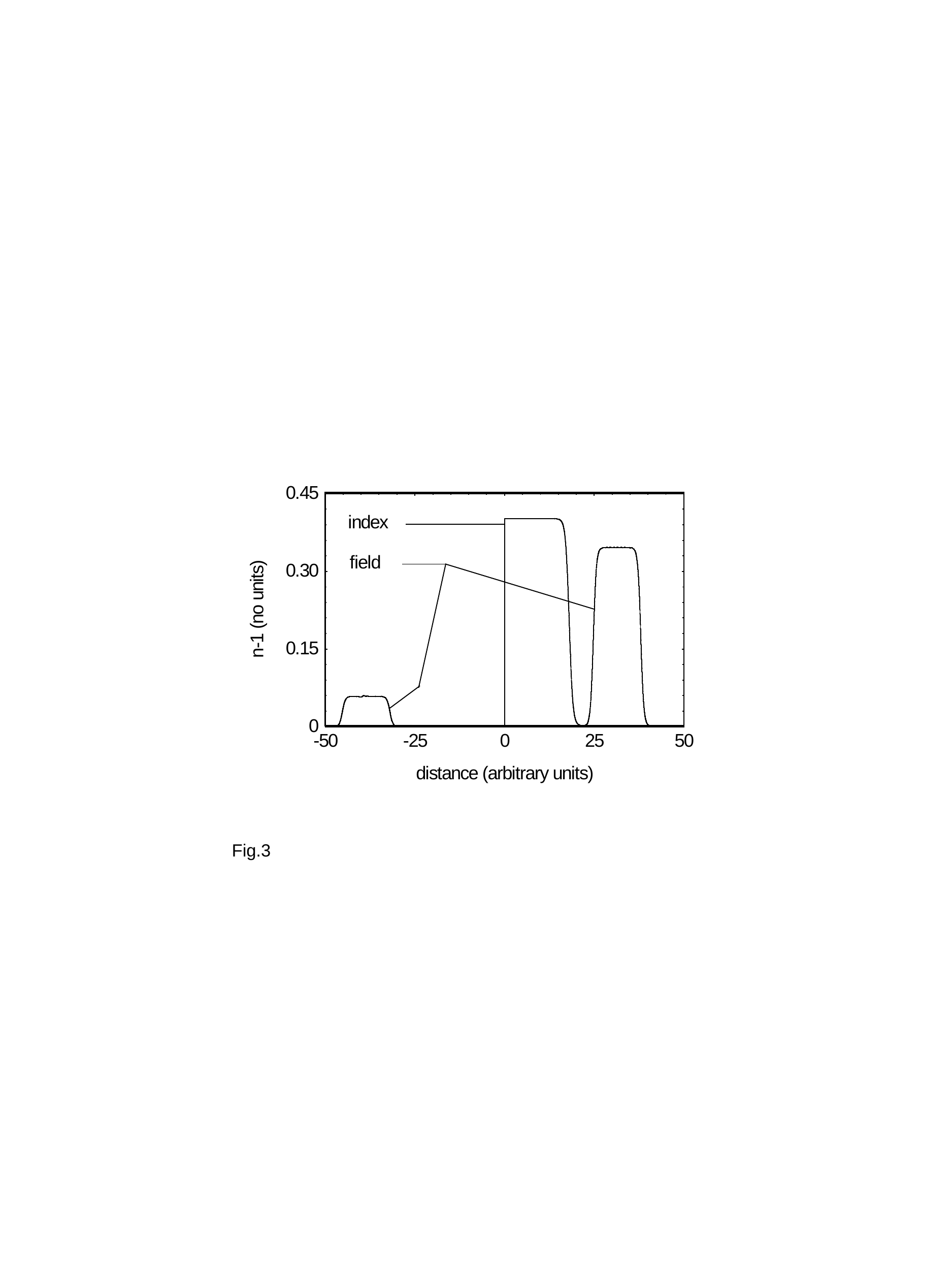}
\caption{Transmitted field has left the medium.}
\label{fig3}
\end{figure}
\par
Conservation of linear momentum cause more problems than conservation of
energy because linear momentum is a directed quantity that changes sign
upon reflection.
Surface reflection takes momentum from the field and transfers the
momentum to the material through radiation pressure.
Once the field has passed entirely through the surface, as in Fig.~2,
there are no more surface forces and the block of material moves with
constant velocity carrying a momentum ${\bf G}_{kinematic}$.
Note that ${\bf G}_{kinematic}$ is not the kinetic momentum described by
Barnett \cite{BIBarn}.
Here, in Fig.~3, all the fields have left the material and have
well-defined momentums in the vacuum.
Then, we can write
\begin{equation}
{\bf G}_i={\bf G}_r+{\bf G}_T+{\bf G}_{kinematic}
\label{EQxxx4.04}
\end{equation}
or
\begin{equation}
\int_{\sigma} \alpha \tilde A_i^2 {\bf\hat e}_z dv =
-\int_{\sigma} \alpha\tilde A_r^2 {\bf\hat e}_z dv+
\int_{\sigma} \alpha\tilde A_T^2 {\bf\hat e}_z dv 
+{\bf G}_{kinematic}
\label{EQxxx4.05}
\end{equation}
by conservation of linear momentum.
Here, $\alpha={\omega_d^2}/(2c^3)$ is a useful combination of
coefficients.
Substituting Eq.~(\ref{EQxxx4.03}) into Eq.~(\ref{EQxxx4.05}) we find
that the kinematic momentum of the material is
\begin{equation}
{\bf G}_{kinematic} = \int_{\sigma}
2\alpha\tilde A_r^2 {\bf\hat e}_z dv
+U_{kinematic}/c \, .
\label{EQxxx4.06}
\end{equation}
Taking $U_{kinematic}/c$ to be negligible, we find that the kinematic
momentum is twice the momentum of the reflected field, but in the
forward direction as determined by the direction of the incident field
such that
\begin{equation}
{\bf G}_{kinematic} =-2{\bf G}_r=
-\int_{\sigma}2\frac{{\bf B}_r\times{\bf \Pi}_r}{c} dv \, .
\label{EQxxx4.07}
\end{equation}
The sign is a bit awkward, but necessary, because 
${\bf B}_r\times{\bf \Pi}_r$ is in the backward (negative) direction
while the kinematic momentum is in the forward direction.
\par
Now we can return to the situation of Fig. 2 with the refracted
pulse entirely within the medium.
There is a linear momentum that is associated with the propagating field
in the medium \cite{BIGord} although its composition in terms of what
portion is field momentum and what portion is material momentum remains
disputed \cite{BIPfei}.
Whatever the composition, the momentum that travels with the field
through the material, ${\bf G}_t$, must be equal to the well-defined
momentum of the field
\begin{equation}
{\bf G}_t(t_1)= {\bf G}_T(t_2)=
\int_{\sigma}\frac{{\bf B}_T\times{\bf \Pi}_T}{c} dv
\label{EQxxx4.08}
\end{equation}
that has exited the material through the antireflection coating.
Applying conservation of energy, Eq.~(\ref{EQxxx3.10}), we find that
${\bf \Pi}_t=\sqrt{n}{\bf \Pi}_T$, ${\bf B}_t=\sqrt{n}{\bf B}_T$, and
\begin{equation}
{\bf G}_t(t_1)=
\int_{\sigma}\frac{{\bf B}_t\times{\bf \Pi}_t}{c} dv
\label{EQxxx4.09}
\end{equation}
is the total momentum that travels with the field inside the medium.
When comparing Eqs.~(\ref{EQxxx4.08}) and (\ref{EQxxx4.09}), recall
that the refracted field is spatially narrower than the transmitted
field.
Substituting Eqs~(\ref{EQxxx4.07})--(\ref{EQxxx4.09}) into
Eq.~(\ref{EQxxx4.04}), we find that the total momentum ${\bf G}_{tot}$
is the sum of well-defined quantities for the refracted momentum, the
reflected momentum, and the kinematic momentum
\begin{equation}
{\bf G}_{tot}=
\int_{\sigma}\frac{{\bf B}_t\times{\bf \Pi}_t}{c} dv
+\int_{\sigma}\frac{{\bf B}_r\times{\bf \Pi}_r}{c} dv
-\int_{\sigma}2\frac{{\bf B}_r\times{\bf \Pi}_r}{c} dv .
\label{EQxxx4.10}
\end{equation}
The identification of ${\bf G}_{tot}$ with ${\bf G}_{i}$ is proven
by demonstrating conservation of the total momentum.
Substituting the definitions of the fields, Eqs.~(\ref{EQxxx2.03}) and
(\ref{EQxxx2.04}), and the Fresnel relations, Eqs.~(\ref{EQxxx3.18}) and
(\ref{EQxxx3.19}), into Eq.~(\ref{EQxxx4.10}) proves that the total
momentum is conserved.
The total momentum has a definite electromagnetic component
\begin{equation}
{\bf G}_{em}=
\int_{\sigma}\frac{{\bf B}_t\times{\bf \Pi}_t}{c} dv
+\int_{\sigma}\frac{{\bf B}_r\times{\bf \Pi}_r}{c} dv
\label{EQxxx4.11}
\end{equation}
that is associated with the propagating field.
The material component
\begin{equation}
{\bf G}_{matl}= 
-\int_{\sigma}2\frac{{\bf B}_r\times{\bf \Pi}_r}{c} dv 
\label{EQxxx4.12}
\end{equation}
is the momentum of the block of dielectric.
In the next section, we will relate the kinematic movement of the block
to the Fresnel surface force.
Now we consider the general case of a dielectric block with index $n_2$
in an inviscid dielectric fluid of index $n_1$.
The momentum of the field in the dielectric block is known to be
\begin{equation}
{\bf G}_{em}=
\int_{\sigma}\frac{{\bf B}\times{\bf \Pi}}{c} dv
\label{EQxxx4.13}
\end{equation}
by Eq.~(\ref{EQxxx4.09}) and by prior work \cite{BIy,BIz,BISPIE,BIx}.
Then repeating the above analysis, the total, electromagnetic, and 
material momentums are still given by the formulas,
Eqs.~(\ref{EQxxx4.10})--(\ref{EQxxx4.12}), although the transmitted and
reflected fields are different in accordance with the Fresnel relations,
Eqs.~(\ref{EQxxx3.18}) and (\ref{EQxxx3.19}).
\par
\section{Electromagnetic Continuity Equations}
\par
The field imparts a surface force to the material due to the change of
sign of the electromagnetic momentum upon reflection.
By Newton's third law, the material accelerates, increasing in momentum.
By Newton's second law, the material imposes an equal force
on the electromagnetic field and momentum is extracted from the field.
Clearly, the subsystems are open systems as momentum is removed from the
field and transferred to the material by the surface force, but the
total system is thermodynamically closed and the total linear momentum,
as well as the total energy, is conserved.
\par
Consider a quasimonochromatic pulse incident on an arbitrarily large
homogeneous medium.
The medium is draped with a gradient-index antireflection coating and
the index changes sufficiently slowly that Helmholtz forces are
negligible.
Then a stationary medium remains stationary, the system is
thermodynamically closed, and energy and momentum are conserved.
The tensor continuity equation is 
\begin{equation}
\bar\partial_{\beta} T^{\alpha\beta}=0 \, ,
\label{EQxxx5.01}
\end{equation}
where
\begin{equation}
\bar\partial_{\beta} =\left ( \frac{n}{c}\frac{\partial}{\partial t}, 
\frac{\partial}{\partial x}, \frac{\partial}{\partial y}, 
\frac{\partial}{\partial z} \right ) 
\label{EQxxx5.02}
\end{equation}
is the material four-divergence
operator \cite{BIy,BIz,BISPIE,BIFinn,BIx,BILag},
\begin{equation}
T= \left [
\begin{matrix}
&({\bf \Pi}^2+{\bf B}^2)/2
&({\bf B}\times{\bf \Pi})_1
&({\bf B}\times{\bf \Pi})_2
&({\bf B}\times{\bf \Pi})_3
\cr
&({\bf B}\times{\bf \Pi})_1
&W_{11}
&W_{12}
&W_{13}
\cr
&({\bf B}\times{\bf \Pi})_2
&W_{21}
&W_{22}
&W_{23}
\cr
&({\bf B}\times{\bf \Pi})_3
&W_{31}
&W_{32}
&W_{33}
\cr
\end{matrix}
\right ]
\label{EQxxx5.03}
\end{equation}
is the total energy--momentum tensor \cite{BIy,BIz,BISPIE,BIx}, and
\begin{equation}
W_{ij}=-\Pi_i\Pi_j-B_iB_j+\frac{1}{2}({\bf \Pi}^2+{\bf B}^2)\delta_{ij} 
\label{EQxxx5.04}
\end{equation}
is the stress tensor \cite{BIy,BIz,BISPIE,BIx}. 
\par
The imposition of a step-index interface on the incident surface of the
solid material is accompanied by a surface force ${\bf F}$ due to
Fresnel reflection.
For a field of cross-sectional area $A$ with square temporal dependence,
\begin{equation}
\Delta{\bf G}_{matl}=
-2A\frac{{\bf B}_r\times{\bf \Pi}_r}{c}\frac{c\Delta t}{n_1}
\label{EQxxx5.05}
\end{equation}
is found by integration of Eq.~(\ref{EQxxx4.07}).
Here, as in the preceding section, $n_1$ is the refractive index of the
region from which the field originates, that is, the index of the
dielectric fluid (or vacuum $n_1=1$) in which the dielectric block is
immersed.
Then
\begin{equation}
{\bf F}=\frac{n_1}{c}\frac{\Delta{\bf G}_{matl}}{\Delta t}=
-2A\frac{{\bf B}_r\times{\bf \Pi}_r}{c} \,.
\label{EQxxx5.06}
\end{equation}
The force must represent a source or sink of electromagnetic momentum
and it must therefore have the same time dependence as the momentum
continuity equation.
For a field in the plane-wave limit that is normally incident on the
block of material, the radiation pressure
\begin{equation}
\frac{{\bf F}}{A}=\frac{1}{A}\frac{n_1}{c}
\frac{c\Delta {\bf G}_{matl}}{\Delta t}
= -2{\bf B}_r\times{\bf \Pi}_r 
\label{EQxxx5.07}
\end{equation}
acts on the incident surface at $z=0$.
Then the radiation pressure can be represented in terms of a force
density as
\begin{equation}
{\bf f} = (-2{\bf B}_r\times{\bf \Pi}_r) \delta(z) \, .
\label{EQxxx5.08}
\end{equation}
There is no source or sink of electromagnetic energy 
so we can write a four-force density
\begin{equation}
f_{\alpha}=\left (0,(-2{\bf B}_r\times{\bf \Pi}_r)\delta (z)\right ) \,.
\label{EQxxx5.09}
\end{equation}
Then the tensor continuity equation for the unimpeded flow of the
electromagnetic field, Eq.~(\ref{EQxxx5.01}), becomes 
\begin{equation}
\bar\partial_{\beta} T^{\alpha\beta}=f_{\alpha} 
\label{EQxxx5.10}
\end{equation}
for a piecewise homogeneous medium.
Because the force density is a sink of the electromagnetic momentum
density, there is an equal and opposite force that acts as a source for
the kinematic momentum of the material.
Using the results of Ref. \cite{BILag} we can derive Newton's second law
$$
{\bf F}=M\frac{d{\bf v}}{d(t/n_1)}
$$
for a material body of mass $M$ immersed in a dielectric fluid of index
$n_1$.
Then the momentum conservation law for the solid block of material is
\begin{equation}
{\bf F}= M\frac{n_1}{c} \frac{cd{\bf v}}{dt}=
\int_{\sigma} -2{\bf B}_r\times{\bf \Pi}_r \delta (z)dv \,.
\label{EQxxx5.11}
\end{equation}
Note that the tensor continuity equation for a flow of non-interacting
material particles that is based on a dust tensor that is used in
Refs.~\cite{BIPfei} and \cite{BIObuk}, for example, does not apply here
because we have posited a solid dielectric.
The components of the tensor continuity equation, Eq.~(\ref{EQxxx5.10}),
are the energy continuity equation,
\begin{equation}
\frac{n}{c}\frac{\partial}{\partial t}
\left ( \frac{1}{2} \left ( {\bf \Pi}^2+{\bf B}^2 \right ) \right )
+\nabla \cdot\frac{{\bf B}\times{\bf \Pi}}{c}=0\, ,
\label{EQxxx5.12}
\end{equation}
and the momentum continuity equation,
\begin{equation}
\frac{n}{c}\frac{\partial}{\partial t}
({\bf B}\times{\bf \Pi})+ \nabla \cdot W=
(-2{\bf B}_r\times{\bf \Pi}_r) \delta (z) \,.
\label{EQxxx5.13}
\end{equation}
The momentum continuity equation, Eq.~(\ref{EQxxx5.13}), explicitly
displays the relation between the change in momentum on the left-hand
side and the effect of the force, acting as a momentum sink, on the
right-hand side.
The momentum continuity equation is not an exact composition of 
Eqs.~(\ref{EQxxx2.12})--(\ref{EQxxx2.15}) because boundary conditions
impose additional constraints.
\par
One of the enduring questions of the Abraham--Minkowski controversy is
why the Minkowski momentum is so often measured experimentally while
the Abraham form of momentum seems to be so favored in theoretical work.
We now have the tools to answer that question.
The Minkowski momentum is not measured directly, but inferred from a
measured index dependence of the optical force on a mirror placed in a
dielectric fluid \cite{BIPfei,BIBarn,BIExp}.
Because the field is completely reflected at the mirror, the force
on the mirror is
\begin{equation}
{\bf F} = \frac{n}{c}\frac{d}{dt}(2c{\bf G})
=\frac{n}{c}\frac{d}{dt}\int_V
2{\bf B}\times{\bf \Pi} \delta (z)dv \, .
\label{EQxxx5.14}
\end{equation}
The measured force on the mirror is directly proportional to the
refractive index $n=n_1$ of the fluid \cite{BIPfei,BIExp}.
On the other hand, if we were to assume ${\bf F}=d{\bf G}/dt$, then we
can write Eq.~(\ref{EQxxx5.14}) as
\begin{equation}
{\bf F} = \frac{1}{c}\frac{d}{dt}\int_V
{2{\bf D}\times{\bf B}}\delta (z) dv
\label{EQxxx5.15}
\end{equation}
using ${\bf D}=-n{\bf \Pi}$.
Then one might infer that the momentum of the field in the dielectric
fluid is the Minkowski momentum.
Instead, we see that the electromagnetic momentum that is obtained from
an experiment that measures the optical force on a mirror depends on the
theory that is used to interpret the results.
However, based on the changes to continuum electrodynamics that are
necessitated by conservation of energy and momentum by the propagation
of light in a continuous medium, we find that Eq.~(\ref{EQxxx5.14}) is
the correct relation between the force on the mirror and the 
momentum of the field in a dielectric.
\par
\section{Conclusion}
\par
The extraordinary persistence of theoretical and
experimental inconsistencies surrounding the Abraham--Minkoswski
controversy \cite{BIPfei,BIMilBoy,BIKemp,BIBaxL,BIBarL,BIMan,BIGriffRL}
regarding the energy--momentum tensor for light in a linear medium
suggested the need to re-examine the role of conservation of energy
and conservation of momentum in classical continuum electrodynamics.
We found that it is necessary to give up the classical macroscopic
Maxwell equations in order to preserve the tensor form of the energy and
momentum conservation laws \cite{BIy,BIz,BISPIE,BIx,BIf}.
Then we must re-examine the body of work that has been built upon the
historical forms of the macroscopic Maxwell equations.
In this article, we derived equations of motion, boundary conditions,
continuity equations and radiation forces for the limiting case of
macroscopic fields ${\bf B}$ and ${\bf \Pi}$ propagating through a
piecewise-homogeneous linear dielectric medium.
\par

\end{document}